# Trap-limited electrical properties of organic semiconductor devices

*Donghyun Ko[1], Gyuhyeon Lee[2], Kyu-Myung Lee[2], Yongsup Park[2,3*], and Jaesang Lee[1*]*

[1]Department of Electrical and Computer Engineering, Inter-University Semiconductor Research Center, Seoul National University, Seoul, 08826, Republic of Korea

[2]Department of Physics and Research Institute of Basic Sciences, Kyung Hee University, Seoul, 02447, Republic of Korea

[3]Department of Information Display, Kyung Hee University, Seoul, 02447, Republic of Korea

[*]Email: jsanglee@snu.ac.kr; parky@khu.ac.kr

**Abstract**

We investigated the electrical properties of a unipolar organic device with traps that were intentionally inserted into a particular position in the device. Depending on their inserted position, the traps significantly alter the charge distribution and the resulting electric field as well as the charge transport behavior in the device. In particular, as the traps are situated closer to a charge-injection electrode, the band bending of a trap-containing organic layer occurs more strongly so that it effectively imposes a higher charge injection barrier. We propose an electrical model that fully accounts for the observed change in the electrical properties of the device with respect to the trap position.

**Introduction**

Traps are initially present or can be generated in any type of amorphous semiconductor device partly or entirely based on organic molecules, such as organic light-emitting diodes, organic photovoltaics, organic thin-film transistors and even perovskite and quantum-dot-based devices [1-8]. Hence, it has long been of particular interest to researchers to understand the impact of traps because not only is their presence the inevitable nature of organics, but they also critically affect the device characteristics. For example, traps can deteriorate the optical performance of organic optoelectronic devices in which they quench excitons or localize free charges which then nonradiatively recombine with the opposite charges [6,8-10]. In addition, the required driving voltage must increase to supply additional charges into the device with traps for maintaining the same level of a current, leading to a higher device resistivity.

Traps can be assumed to exhibit Gaussian density of states (DOS) due to the disordered arrangement of the molecules consisting of organic devices [11-13]. With respect to the density and energetic depth and width of the Gaussian traps, several studies examined how the current density vs. voltage (*J-V*) characteristics of unipolar organic devices are determined [14,15]. However, the *spatial disorder of traps* and its potential impact on the device characteristics have not been thoroughly discussed in most previous studies. Note the strong position dependence of degradation reactions, e.g., the ion migration from a metal electrode into organics [16-18] or the molecular bond dissociation induced by the highly excited states that are in general nonuniformly generated in organic optoelectronic devices [6,19-22]. That is, the spatial distribution of such degradation-induced traps is highly likely nonuniform, whereby they affect the charge and exciton dynamics and hence the device characteristics in a complicated manner, which has been rarely discussed so far. The primary reason may lie in the extreme difficulty of identifying the specific position of a variety of traps with unknown energetics and low concentrations (<1wt%) and the decoupling of their mixed effects.

In this study, we investigate the effects of a trap position on the *J-V* characteristics of a hole-only unipolar organic device (HOD). The HOD includes a 200 nm-thick hole transport layer (HTL), of which a 5 nm-thick slab *at a particular position* is doped with molecules with *known energetics* that can act as hole traps. We found that the distribution of holes and the associated electric field profile for the trap-containing HODs markedly differ with respect to the trap position, resulting in their distinct *J-V* characteristics. In particular, as the traps are inserted closer to a hole-injection electrode (i.e., an anode), the holes localized by the traps more strongly induce the band bending of the HTL so that hole injection is further restricted. We considered such trap-induced band bending to obtain the boundary conditions (B.C.) for the hole densities and electric potential for a *J-V* model, with which we fully account for the observed electrical characteristics of the HOD by varying the trap position.

**Theory**

The trap-limited *J-V* model is formulated according to the classical- and organic-semiconductor theories, including the drift-diffusion current model, Poisson's equation, and Fermi-Dirac statistics taking into consideration the Gaussian DOS for both hole-transport and -trapping states. The model is designed to fit the *J-V* characteristics of the trap-containing HODs, for which it calculates an electrical potential ($V$), free hole density ($p$), and trapped hole density ($p_t$) as functions of a position ($x$) in the HTL with a thickness of $d = 200$ nm. The hole traps are assumed to be situated in the range of [$x_t - 2.5$ nm, $x_t + 2.5$ nm] within a 5 nm-thick slab, where $x_t$ is the center of the slab (see Supplemental Material for the detail of the model [23]).

To solve the second-order differential equations for $V(x)$ and $p(x)$ in the model, we should define the B.C. for these two variables at $x = 0$ and $x = d$ at which hole-injection and -extraction occur, respectively. The B.C. for $V$ are given by $V(0) = V_a - V_{bi}$ and $V(d) = 0$, where $V_a$ is the applied voltage and $V_{bi}$ is the built-in potential. Here, $V_{bi}$ can be viewed as the difference between the energy barriers for hole injection at $x = 0$ and extraction at $x = d$, denoted by $\phi_{inj}$ and $\phi_{ext}$, respectively, at a thermal equilibrium of $V_a = 0$. Namely,

$$V_{bi} = \phi_{ext} - \phi_{inj}(x_t, V_a = 0), \text{ where} \tag{1a}$$

$$\phi_{inj}(x_t, V_a) = \Delta_0 + \Delta(x_t) - l\sqrt{V_a}. \tag{1b}$$

**Eq. (1b)** indicates that the hole-injection barrier ($\phi_{inj}$) can vary with respect to $x_t$ and $V_a$, as schematically described in **Fig. 1**. Here, $\Delta_0$ in **Fig. 1(a)** represents the *intrinsic* band bending at equilibrium, which naturally arises if the HTL is in contact with a high workfunction anode [24]. This can be understood as follows: in an instant after contact formation, the substantial density of holes diffuses from the anode into the HTL to align the Fermi level [12,13,25-27]. Then, the accumulated holes near the anode/HTL interface create an electric field towards the anode, resulting in the downward curvature of an electric potential (i.e., $\Delta_0$). Note that the high workfunction anode can render an *Ohmic-like* contact for efficient hole injection at the anode/HTL junction rather than a Schottky contact (*vide infra*).

Now, in the presence of traps at $x = x_t$ *near the anode* as shown in **Fig. 1(b)**, the HTL band curves further downwards by $\Delta(x_t)$ due to the additional electric field that is exerted towards the anode by the trapped holes. On the other hand, the application of $V_a > 0$ opposes the intrinsic ($\Delta_0$) and *extrinsic* ($\Delta(x_t)$) band bending so that $\phi_{inj}$ is effectively reduced with respect to that at equilibrium (i.e., $V_a = 0$) (see **Fig. 1(c)**). We assumed that this bias-dependent barrier-lowering effect depends on the square root of $V_a$ in **Eq. (1b)** inspired by the thermionic emission current model, where the current density is given by $J \propto \exp\left[-\left(\frac{\phi_{inj} - \sqrt{qF/4\pi\varepsilon}}{kT}\right)\right]$ with $q$ the elementary charge, $k$ the Boltzmann constant, $\varepsilon$ the permittivity and $T$ the temperature. Here, $\phi_{inj}$ is lowered by $\sqrt{q/4\pi\varepsilon d}\sqrt{V_a}$ with an assumption of the constant electric field of $F = V_a/d$ across the organic layer [11,28,29]. However, considering the space charge effect in our HOD, we replaced the constant prefactor $\left(\sqrt{q/4\pi\varepsilon d}\right)$ with a scale factor of $l$ that is set as a free parameter in our $J$-$V$ model.

The B.C. for $p$ can be given according to the Fermi-Dirac statistics for holes and the Gaussian DOS for the HTL as:

$$p(x = 0, d) = \int_{-\infty}^{\infty} \frac{N_H}{\sigma_H \sqrt{2\pi}} \exp\left[-\frac{(E - E_0)^2}{2\sigma_H^2}\right] \frac{1}{1 + \exp[(\phi_{inj,ext} - E)/kT]} dE, \qquad (2)$$

where $E$ is an electronic state energy for holes, $N_H$ is the total molecular density, $E_0$ is the center of the Gaussian DOS, and $\sigma_H$ is the Gaussian width for the HTL molecules. **Eq. (2)** indicates that the *injected* hole density ($p(0)$) is a function of $\phi_{inj}$ and hence, it depends on $x_t$ and $V_a$. For example, extrinsic band bending ($\Delta(x_t)$) caused by traps enhances the injection barrier ($\phi_{inj}$) and thus reduces $p(0)$, whereas a positive $V_a$ lowers $\phi_{inj}$, thereby increasing $p(0)$. $p(0)$ determines the density of the free holes available in the HTL and hence the conductivity of the HODs according to the drift-diffusion equation.

**Experiment**

The *inverted* structure for the HODs is as follows: indium-tin oxide (ITO) / 10 nm MoO$_3$ / 200 nm tris(4-carbazoyl-9-ylphenyl)amine (TCTA) / 3 nm 4,4'-bis(N-carbazolyl)biphenyl (CBP) / 10 nm MoO$_3$ / 100 nm Al. Here, an anode with the sequence of Al, MoO$_3$, and CBP forms an *Ohmic-like* hole-injection contact with the TCTA HTL due to a CBP interlayer with a high ionization energy [30]. This enables us to mainly focus on the hole transport properties in the HTL bulk, while minimizing the injection-limited effects on the $J$-$V$ characteristics. We chose 4,4',4"-tris[2-naphthyl(phenyl)amino]triphenylamine (2-TNATA) as a hole trap molecule due to its shallower highest occupied molecular orbital (HOMO) of −5.3 ± 0.1 eV vs. −5.8 ± 0.1 eV for the TCTA HTL. This 0.5 eV difference eliminates the energetic overlap between the HTL and trap HOMOs, even considering the width of a Gaussian DOS ~ 0.1 eV for typical organics [27,30].

To minimize any *percolating* movement of localized holes through the trap molecules, we render the intermolecular distance between the trap molecules as far as possible by keeping their concentration low ≤ 1 wt% [31,32]. Hence, 2-TNATA was doped at 1 wt% in a 5 nm-thick slab of the 200 nm-thick TCTA HTL, forming a *trap-containing slab* centered at $x = x_t$ within the HTL. The "insertion" of the trap-containing slab at a desired position ($x_t$) was done by the sequential deposition of the partial HTL, slab and then another partial HTL with a precisely controlled thickness. The vacuum pressure in the deposition chamber is kept below ~5×10$^{-7}$ Torr to minimize the introduction of environmental impurities such as oxygen and water molecules into the HODs. The trap depth for 2-TNANA *in* TCTA was measured to be 0.5 ± 0.1 eV with ultraviolet photoelectron spectroscopy (UPS), precisely matching with the difference between their HOMO levels measured independently (see Supplemental Material [23]).

We classified the HODs into two groups as shown in **Fig. 2(a)**. The devices in "group 1" include a trap-containing slab centered at $x_t$ = 0.2$d$ up to 0.9$d$ with an interval of 0.1$d$, relatively far away from the anode. For the devices in "group 2", on the other hand, the slab is situated at a shorter distance of $x_t$ = 0.025$d$ to 0.1$d$ to the anode with a shorter interval of 0.025$d$. "Group 2" was used to investigate strong trap-induced bend bending near the anode with high precision.

**Results and Discussion**

**Figs. 2(b)** and **2(c)** show the *J-V* characteristics of a few select HODs from "group 1" and "group 2", respectively, along with that of a control device without traps (symbols: experimental data, solid and dotted lines: model fits). The control device clearly exhibits a *trap-free* charge transport behavior, consistent with the Mott-Gurney law (i.e., $J \propto V^2$), in contrast to the other devices in "group 1" and "group 2" [33]. This indicates that the *intrinsic* or environmental traps that may be present in all HODs do not act as hole traps [34]. Thus, we confirmed that the only kind of hole traps active in the HODs is the *intentionally* introduced 2-TNATA molecule.

For the "group 1" HODs, the *J-V* model with an assumption of $\Delta(x_t)$ = 0 shows a good agreement with the measured *J-V* data [**Fig. 2(b)**]. This indicates that the trap-induced band bending may be insignificant for the "group 1" HODs whose traps are situated relatively far from their anode (0.2$d$ ≤ $x_t$ ≤ 0.9$d$). However, for the "group 2" HODs with the traps in proximity to the anode with $x_t$ ≲ 0.1$d$, the *J-V* model with the same assumption (i.e., $\Delta(x_t)$ = 0) largely *overestimates* the actual *J-V* trend or the *conductivity* of the devices [dotted lines in **Fig. 2(c)**]. Furthermore, such a discrepancy progressively increases with decreasing $x_t$. This model overestimation can be resolved by considering the mechanisms that can lessen the device conductivity – (i) the reduced hole mobility of the HTL or (ii) restricted hole injection into the HTL as $x_t$ approaches the anode. While the former can be ruled out as it is not the case for "group 1", the latter can result from *nonvanishing* trap-induced band bending ($\Delta(x_t) \neq 0$ eV) [**Eq. (1b)**]. That is, the nonzero $\Delta(x_t)$ increases $\phi_{inj}$ and hence concomitantly reduces $p(0)$, rendering the HODs to become *resistive*. The *J-V* model with the *revised* $\Delta(x_t)$ values can precisely fit the data for the "group 2" HODs [solid lines in **Fig. 2(c)**].

Note that $\Delta(x_t)$ increases with decreasing $x_t$ only if the traps are in close proximity to the anode ($x_t \lesssim 0.1d$), but otherwise $\Delta(x_t) \approx 0$ eV ($x_t \geq 0.2d$). This is because the trapped holes nearer the anode exert a stronger electric field towards the anode, further inducing downwards band bending (see Supplemental Material for verification [23,35]). Note also that because the density of trapped holes ($p_t$) increase with the applied electric field or voltage ($V_a$), so does the nonzero $\Delta(x_t)$ for $x_t \lesssim 0.1d$, which arises due to $p_t$ in the first place. However, the variation of $p_t$ vs. $V_a$ is small despite the addition of free holes ($p$) due to the much larger density of $p_t > 10^{24}$ cm$^{-3}$ vs. $p \approx 10^{18} \sim 10^{23}$ cm$^{-3}$ across the HODs (*vide infra*), which is attributed to the favorable energy level for the trap states (i.e., $E_t = 0.5 \pm 0.1$ eV) to readily capture a majority of free holes even at a very low $V_a$. This leads to the negligible $V_a$-dependence of $\Delta(x_t)$.

On the other hand, the band bending can be mitigated by a positive applied bias ($V_a > 0$) (see **Fig. 1(c)** and **Eq. (1b)**). To account for this bias-dependent barrier lowering for HODs, the fitting parameter $l$ in **Eq. (1b)** was obtained to be 0.09, which is nearly twice the prefactor $\sqrt{q/4\pi\varepsilon d} \approx 0.05$ for the thermionic emission model with the assumption of a constant electric field. This discrepancy may attribute to the *space charge effect* in our HODs, one of the fundamental assumptions in our model (see Supplemental Material [23]) that leads to the nonuniform electric field across the devices (*vide infra*). Another possible explanation can be given in terms of an energy broadening effect in the organic layer at a metal-organic interface. The energy broadening originates from the disorder of interfacial dipoles formed between the metal and organic [11]; however, upon the application of the positive bias, such *disordered* dipoles would tend to align in the direction of the electric field. This, as a result, can reduce the energy broadening of the organic layer with a *narrowed* Gaussian DOS such that charge injection from the metal into the organic layer may be facilitated.

To investigate the overall trend of the J-V characteristics with respect to the trap position, we present a contour map of log J as functions of $x_t$ and $V_a$ in **Fig. 3(a)**. The black contour line at $J_0 = 0.01$ mA/cm$^2$ shows two trends: (1) the required $V_a$ to attain $J_0$ increases with decreasing $x_t$ from $0.9d$ to $0.2d$ (i.e., the HODs become resistive); (2) for $x_t \lesssim 0.2d$, the trend is reversed (i.e., the HODs become conductive). To understand this, we plot in **Fig. 3(b)** the distributions of free ($p$) and trapped holes ($p_t$) along with the electric field profile ($F$) at $J_0 = 0.01$ mA/cm$^2$ for the three selected HODs with traps inserted at $x_t = 0.05d$, $0.3d$, and $0.5d$, respectively. The commonality for these devices is that $p$ sharply drops near the trap-inserted position at $x = x_t$ due to strong charge localization, manifested also as a large local density of $p_t$ therein. In the case of $x_t = 0.5d$, for example, the proportion of trapped holes vs. total holes, $\int_0^d p_t(x)dx / \int_0^d [p(x) + p_t(x)]dx$, is approximately 40%. This is remarkable in that only 1 wt% of the traps within a 5 nm-thick slab (i.e., 2.5% of the entire thickness) can localize nearly a half of the total available holes. It is the result of the trap depth of $E_t = 0.5 \pm 0.1$ eV being deep enough to readily and exothermically attract holes from the transport energy level.

The remaining free holes ($p$) between the trap-containing slab at $x = x_t$ and the cathode at $x = d$ are nearly negligible, for which $p$ in this region is smaller by at least ~3 orders of magnitude than the injected hole density ($p(0)$). As a result, the electric field in the direction of the cathode ($F > 0$) surges at which a large local $p_t$ arises near $x = x_t$, but it flattens out in the region of $x_t \lesssim x \leq d$ due to the scarcity of overall holes therein according to $F(x) = F(0) + \int_0^x \frac{q}{\varepsilon}[p(x') + p_t(x')]dx'$ [**Fig. 3(b)**]. Therefore, as $x_t$ approaches more the anode, the width of this flattened or *constant-electric field* region at $x_t \lesssim x \leq d$ becomes wider. This leads to the higher $V_a$ to maintain a given $J_0$ according to $V_a \approx -\int_d^0 F(x')dx'$, indicating that the HODs become more resistive with decreasing $x_t$, as is found in trend (1). In short, the free holes available in the devices become more deficient as they are captured by the traps nearer to the anode and this leads to the reduced device conductivity, i.e., requiring the higher $V_a$ to attain $J_0$, as indicated by the decreasing slope of the $J$-$V$ curves for the "group 1" HODs with decreasing $x_t$ [**Fig. 2(b)**].

While the same effect also holds true for the case with $x_t \lesssim 0.2d$, it is overwhelmed by a *counterbalancing* factor that renders the HODs to become conductive with decreasing $x_t$, i.e., trend (2) – an entire negative shift of the $F(x)$ that occurs with decreasing $x_t$ such that the magnitude of $F(x) > 0$ in the direction of the cathode progressively reduces (i.e., at $x_t \lesssim x \leq d$), whereas $F(x) < 0$ in the direction of the anode is further intensified (i.e., at $0 \leq x \lesssim x_t$). This can be understood in terms of the relative distance between the trapped charges ($p_t(x_t)$) and the respective electrodes, which modulates the magnitude of $F(x)$ according to *the method of images*. The observed relationship of $F(x)$ vs. $x_t$ can be more clearly shown by solving the Poisson's equation with $p_t(x_t)$ with a fixed density, excluding the influence of the free charges (i.e., $p(x) = 0$) without loss of generality (see Supplemental Material [23]).

In summary, the turning point of the device conductivity curve (i.e., the contour line at $J_0$) is determined by the interplay of three factors – (i) the broadening of the width of the constant-$F$ region, (ii) the entire negative shift of the $F(x)$ profile and (iii) the trap-induced bend bending ($\Delta(x_t)$). The factors (i) and (iii) render the HODs resistive, whereas the factor (ii) exerts the opposite effect. **Fig. 3(a)** shows the relative contribution of the individual factors to the device conductivity with respect to $x_t$. The factor (i), represented by a dashed line, is dominant in the range of $0.2d \lesssim x_t \leq 0.9d$, whereas the factor (ii) and (iii) become activated with decreasing $x_t \lesssim 0.2d$; however, even though the effect of the factor (iii), rendering the HODs resistive, is present (see the difference between black and red solid lines assuming $\Delta(x_t) > 0$ and $\Delta(x_t) = 0$, respectively), the factor (ii), represented by a dotted line, is so dominant that the HODs become conductive after all, manifested as the turning point of the contour line in **Fig. 3(a)**.

## Conclusions

We investigated the effects of traps on the electrical characteristics of unipolar organic devices by varying the position of the traps in the devices. We found that the traps near an anode induce strong band bending of the organic layer, thereby hindering charge injection into the device. In addition, the position of the traps critically determines the distribution of the free and trapped charges, the electric field, and hence the conductivity of the device. Note that our analysis on organic devices could be easily transferrable to any type of device based on amorphous semiconductors. This is because although the material composition for amorphous semiconductor devices is different, the governing principle of charge dynamics, electronic states arising due to the structural disorder, and their variation due to the presence of traps can be common in a broad sense. Our findings indicate that for amorphous semiconductor devices in which environmental or degradation-induced traps are normally found, a slight spatial disorder of the traps even with a minute amount can largely modulate the entire electrical properties of the devices. It is therefore imperative to figure out what kinds of traps are present in devices in the first place and then the effects of such traps with their originating position and distribution, as was proposed in this study.

## Acknowledgements

The authors acknowledge the financial support from Samsung Display, the Industrial Strategic Technology Development program funded by the Ministry of Trade, Industry & Energy (MOTIE, Korea) (20011059), the National Research Foundation of Korea (NRF) grant funded by the Korean government (MSIT) (2020R1C1C1008659 and 2021R1A2C1012754), and the New Faculty Startup Fund from Seoul National University. The UPS data were measured using the equipment in Multidimensional Materials Research Center at Kyung Hee University (2021R1A6C101A437).

**Figures and captions**

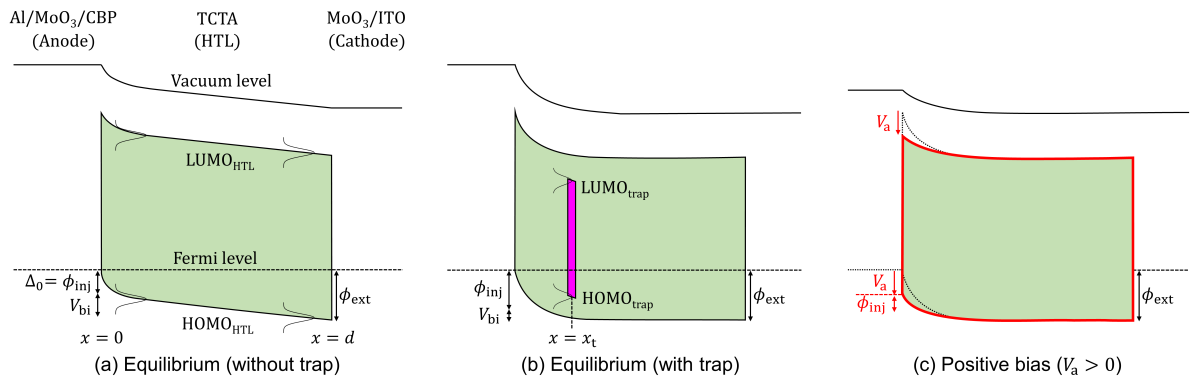

**FIG. 1.** Energy band diagrams for the HODs (a) at thermal equilibrium where *intrinsic* band bending ($\Delta_0$) arises for the Fermi level alignment, effectively acting as a hole-injection barrier ($\phi_{\text{inj}}$); (b) at equilibrium where additional *extrinsic* band bending occurs due to traps inserted near an anode at $x = x_t$, further increasing $\phi_{\text{inj}}$; (c) upon the application of a positive voltage ($V_a > 0$) where band bending is mitigated, reducing $\phi_{\text{inj}}$.

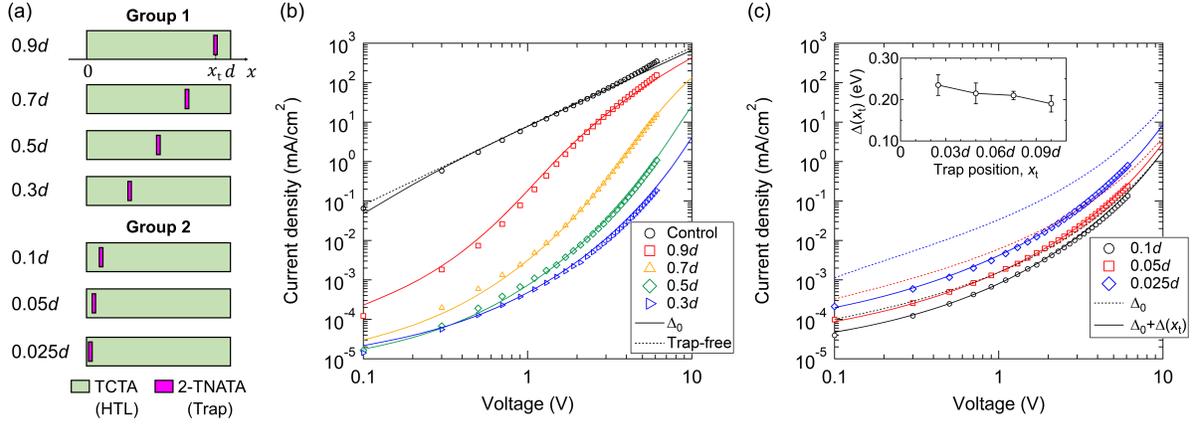

**FIG. 2.** (a) A few select HODs with traps inserted at different positions ($x_t$). The "group 1" and "group 2" HODs have the traps relatively far from ($0.2d \leq x_t \leq 0.9d$), and near to ($x_t \leq 0.1d$), their anode, respectively, with $d$ the thickness of the HTL. (b) The $J$-$V$ characteristics for the "group 1" HODs and the control device without traps (dotted line: $J$-$V$ model following the Mott-Gurney law; solid lines: our $J$-$V$ model only assuming intrinsic band bending ($\Delta_0$); symbols: experimental data). (c) The $J$-$V$ characteristics for the "group 2" HODs (dotted and solid lines: our $J$-$V$ models with the assumptions of $\Delta(x_t) = 0$ and $\Delta(x_t) > 0$, respectively). The inset shows the non-zero $\Delta(x_t)$ values with respect to $x_t$ for the "group 2" HODs.

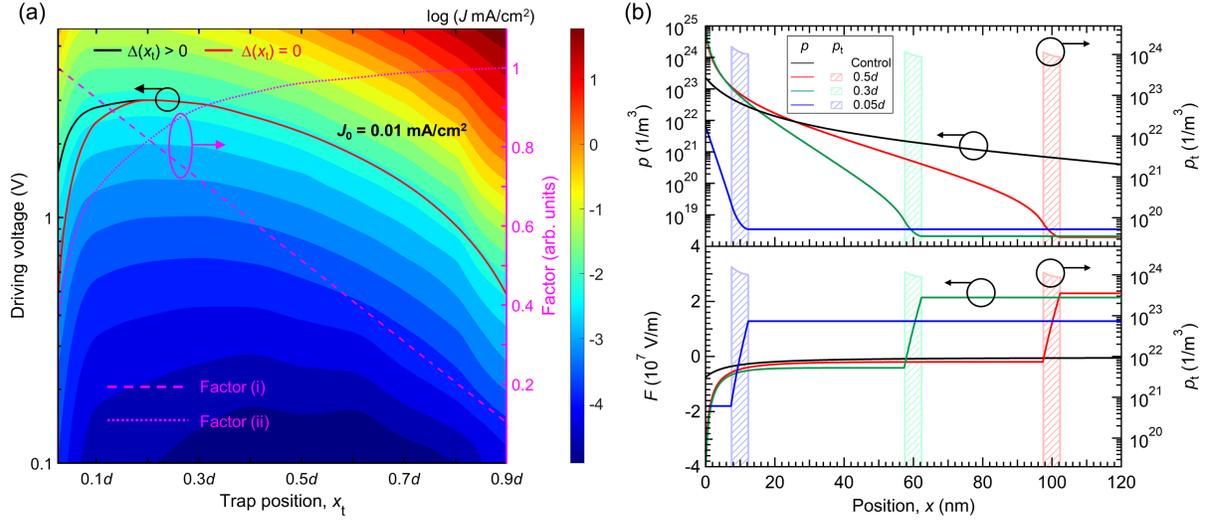

**FIG. 3.** (a) Contour plot of the current density of the HODs with respect to trap position ($x_t$) and driving voltage ($V_a$). Solid lines represent a required $V_a$ to attain $J_0 = 0.01$ mA/cm² for the HODs with traps at $x = x_t$, calculated using the $J$-$V$ model with and without considering the effect of $\Delta(x_t)$. Dashed and dotted lines represent the relative contributions of the width of the constant-$F$ region (factor (i)) and the negative shift of $F(x)$ (factor (ii)), respectively, to the contour line with respect to $x_t$. (b) [upper panel] The profile of the free hole density ($p$; solid line), trapped hole density ($p_t$; diagonal pattern) and [lower panel] electric field ($F$; solid line) for a few select HODs with traps inserted at $x_t = 0.05d$, $0.3d$, and $0.5d$, all of which are calculated at $J_0 = 0.01$ mA/cm².

# Supplemental Material

# Trap-limited electrical properties of organic semiconductor devices

*Donghyun Ko[1], Gyuhyeon Lee[2], Kyu-Myung Lee[2], Yongsup Park[2,3]\*, and Jaesang Lee[1]\**

[1]Department of Electrical and Computer Engineering, Inter-University Semiconductor Research Center, Seoul National University, Seoul, 08826, Republic of Korea

[2]Department of Physics and Research Institute of Basic Sciences, Kyung Hee University, Seoul, 02447, Republic of Korea

[3]Department of Information Display, Kyung Hee University, Seoul, 02447, Republic of Korea

\*Email: jsanglee@snu.ac.kr; parky@khu.ac.kr

## Sec. 1. The complete structure of the *J-V* model

The HODs used in this study operate based on the dynamics of unipolar charges (i.e., holes) associated with traps, which lack charge recombination and excitonic processes. The absence of electrons in the devices is possible by having a sufficiently large energy barrier against electron injection from the cathode (ITO / MoO$_3$) into the TCTA HTL (see **Fig. 1** in the main text). Furthermore, due to the large energy gap of $E_g > 3$ eV for the HTL, the intrinsic electron concentration therein is negligible according to $n_i = N_H \exp(-E_g/2kT)$ with $N_H$ the molecular density, $k$ the Boltzmann constant, and $T$ the temperature.

We assume that a free hole mobility for the HTL, $\mu = 1.4 \times 10^{-4}$ cm$^2$/V·s [1], is low enough so that not only can trapped holes ($p_t$) but also free holes ($p$) be treated as space charge. Hence, using the Poisson's equation, the electric potential ($V$) in the HTL can be described as follows:

$$\frac{d^2 V(x)}{dx^2} = -\frac{q}{\varepsilon}[p(x) + p_t(x)]. \tag{S1}$$

**Eq. (S1)** is coupled with the continuity equation under steady-state conditions, where the current density, *J*, for the HOD is given using the drift-diffusion equation as:

$$\frac{dJ(x)}{dx} = \frac{d}{dx}\left(-q\mu p(x)\frac{dV(x)}{dx} - qD\frac{dp(x)}{dx}\right) = 0, \tag{S2}$$

where *D* is the diffusion coefficient that is assumed to satisfy the Einstein relationship. Note that $\mu$ in **Eq. (S2)** can be assumed to be constant, i.e., independent of the electric field and charge density, without loss of generality for this particular study (see Supplementary Material, **Sec. 2**).

The trapped hole density, $p_t(x)$, in **Eq. (S1)** can be given by integrating the product of the Gaussian DOS for traps and the Fermi-Dirac occupancy for holes over the electronic state energy, *E*, as follows:

$$p_t(x) = S(x)\int_{-\infty}^{\infty} \frac{N_t}{\sigma_t \sqrt{2\pi}} \exp\left[-\frac{(E-E_t)^2}{2\sigma_t^2}\right] \frac{1}{1+\exp[\{E_{FH}(x) - E\}/kT]} dE, \tag{S3}$$

where $N_t$, $E_t$, and $\sigma_t$ are the total density, Gaussian depth, and width of the trap molecules, respectively. Here, $S(x) = 1$ if *x* belongs to the range of [$x_t - 2.5$ nm, $x_t + 2.5$ nm] and $S(x) = 0$ otherwise. $E_{FH}$ is the quasi-Fermi level with respect to the center of the Gaussian HTL as follows:

$$E_{FH}(x) = \frac{kT}{q}\ln\left[\frac{N_H}{p(x)}\right] + \sigma_H^2/2kT, \tag{S4}$$

where *k* is the Boltzmann constant and *T* is the temperature fixed at 295 K. Here, the first term in **Eq. (S4)** is obtained according to the Boltzmann approximation with the assumption of the *single-level* HTL state, which can be only allowed in the trap-inserted region where the free hole density (*p*) is nearly negligible compared to the total molecular density ($N_H$) as shown in **Fig. (3b)**, i.e., representing a *nondegenerate case*. Then, the second term is added to account for the *Gaussian distribution* of the HTL states elsewhere [2,3]. Note that we treated the HTL and trap molecules as independent systems without any mutual interactions, and hence, their Gaussian electronic states were assumed to retain their own *mean* energy level and variance (i.e., energy depth).

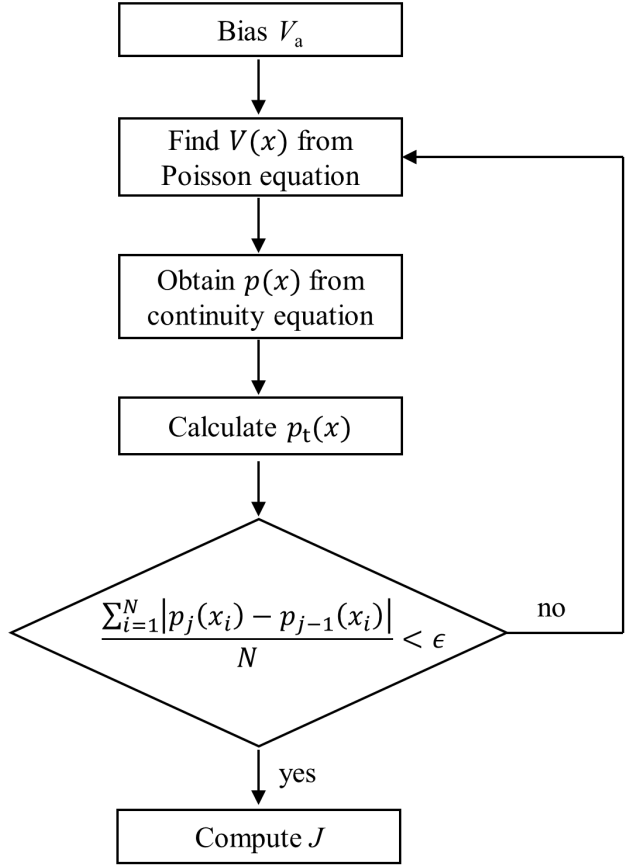

**FIG. S1. The flow chart of the *J-V* model.**

**Fig. S1** shows the flowchart of our *J-V* model. With the given boundary conditions (B.C.) for $p$ and $V$ (see the main text), the **Eqs. (S1)–(S4)** are solved *iteratively* for $p(x)$, until the overall difference in the $p$ values at all given $x$ *between two consecutive loops* becomes smaller than the designated error $\varepsilon$. To be specific, the loop is terminated at a $j$-th trial if the condition $\frac{1}{N}\sum_{i=1}^{N}|p_j(x_i) - p_{j-1}(x_i)| \leq \varepsilon$ is met, where $x_i$ is the $i$-th value of $N$ evenly spaced positions in the HTL between $x = 0$ and $x = d$. Using the *converged* $p(x)$ result, $J$ for a given $V_a$ is computed according to the drift-diffusion current equation, as given in **Eq. (S2)**. The same procedure is repeated for different bias conditions ($V_a$) that yield the corresponding $J$ values, enabling us to obtain complete $J$-$V$ characteristics for the given HODs. The parameters used for the $J$-$V$ model are listed in **Table S1**.

**Table S1. Parameters used in the *J-V* model.**

| Parameter | Value |
|---|---|
| $N_H$ (1/m³) | $5.3 \times 10^{26}$ |
| $\sigma_H$ (eV)[1] | 0.096 |
| $E_0$ (eV) | 0 |
| $E_t$ (eV) | 0.54 |
| $\sigma_t$ (eV) | 0.13 |
| $\phi_{ext}$ (eV) | 0.63 |

## Sec. 2. Gaussian disorder model for a charge carrier mobility

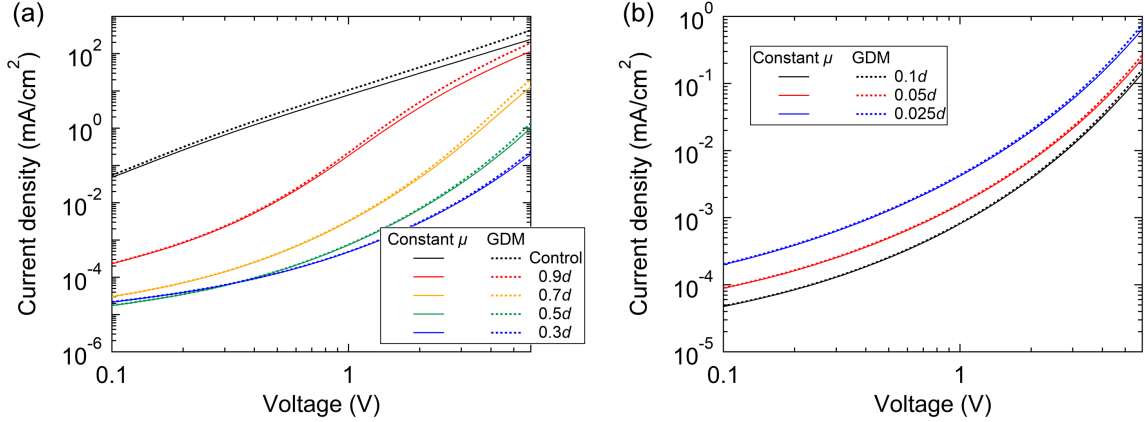

**FIG. S2.** The calculated *J-V* characteristics under the assumptions of the *constant* charge carrier mobility (solid line) and the *Gaussian disorder model (GDM)*-based mobility (dotted line) for (a) the control and "group 1" and (b) for "group 2" HODs.

For disordered organic semiconductors, the charge carrier mobility is known to vary with respect to the electric field ($F$) and charge density ($p$) according to the Gaussian Disorder model (GDM) [4-6]. We investigated the impact of such $F$ and $p$-dependence of the charge carrier mobility on our model and analysis. Following Pasveer *et al.* [4], the hole mobility can be given as:

$$\mu(F(x), p(x)) \approx \mu_0 g_1(F(x)) g_2(p(x)), \tag{S5}$$

where $\mu_0 = 1.4 \times 10^{-4}$ cm$^2$/V·s being the mobility under the condition of a low electric field and a charge density, $g_1(F)$ and $g_2(p)$ represent the *modulating* factors due to the increasing electric field and charge density, respectively. To be specific, $g_1(F)$ and $g_2(p)$ can be given as:

$$g_1(F(x)) = \exp\left\{0.44(\hat{\sigma}^{3/2} - 2.2)\left[\sqrt{1 + 0.8\left(\frac{F(x)qa}{\sigma}\right)^2} - 1\right]\right\} \quad and \tag{S6a}$$

$$g_2(p(x)) = \exp\left[\frac{1}{2}(\hat{\sigma}^2 - \hat{\sigma})(2p(x)a^3)^\delta\right], \tag{S6b}$$

where $\delta = 2\frac{\ln(\hat{\sigma}^2 - \hat{\sigma}) - \ln(\ln 4)}{\hat{\sigma}^2}$, $a = N_\mathrm{H}^{-1/3}$ and $\hat{\sigma} = \sigma/kT$.

By substituting the GDM-based mobility [**Eq. (S5)**] into **Eq. (S2)**, we calculated the *J-V* characteristics of the "group 1" and "group 2" HODs following the procedure in Sec. 1, which were then compared to those assuming the *constant* charge mobility, as shown in **Fig. S2**. As the applied voltage increases, the *J-V* with the GDM-based mobility deviates only slightly from that with the constant mobility due to increasing $p$ and $F$. That is, in the operational voltage regime for our HODs, i.e., $0 \leq V_a \leq 6$ (V), the discrepancy in log $J$ between two cases is $\leq 10\%$ at all $V$, which is less than an experimental error. Therefore, we confirmed that neither does the assumption of the constant mobility significantly affect our analysis nor increase the complexity of the model and computational cost.

## Sec. 3. *in-situ* ultraviolet photoelectron spectroscopy (UPS)

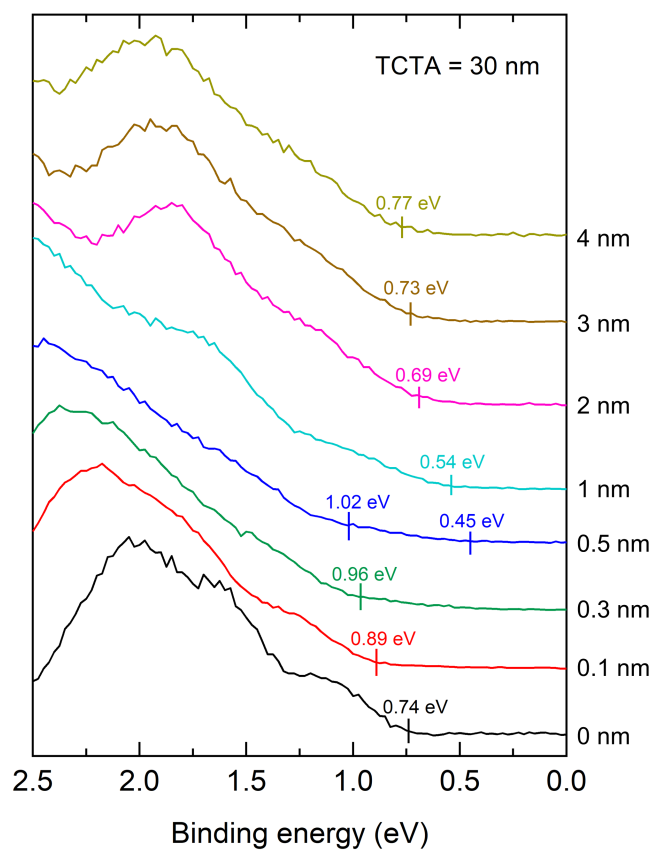

**FIG. S3. UPS spectra of a 2-TNATA film with an increasing thickness grown on a TCTA film.**

The trap depth ($E_t$) used in this study refers to the difference between the HOMO energy levels for the trap (2-TNATA) and HTL (TCTA) molecules. The HOMO energy levels were measured using an *in-situ* UPS method, in which sample deposition and measurement can be performed sequentially in the same vacuum chamber without the need to transfer the sample in the non-vacuum environment between the two processes. The base pressure of the chamber was maintained at ~$10^{-9}$ Torr, and the energy of the He I excitation source was 21.22 eV. The organic layer was deposited directly on the UV-ozone treated ITO substrate. The TCTA HTL was pre-deposited with a fixed thickness of 30 nm, on which the growth of a 2-TNATA film followed with varying thicknesses from 0.1 to 4 nm.

**Fig. S3** shows the UPS spectra of samples with respect to the 2-TNATA film thickness. The distinct onset of the spectrum represents the HOMO with respect to the Fermi level at a binding energy of 0 eV. The signal from the 2-TNATA film emerges and becomes dominant if its thickness is ≥ 0.5 nm; otherwise, only that from the TCTA film is observed. When both signals are present, the difference between the HOMO energy levels for 2-TNATA and TCTA was measured to be $0.5 \pm 0.1$ eV. Hence, the trap depth of $E_t = 0.54$ eV used in the *J-V* model (**Table S1**) agrees well with the measured HOMO offset between the traps (2-TNATA) and the HTL (TCTA) according to the UPS data.

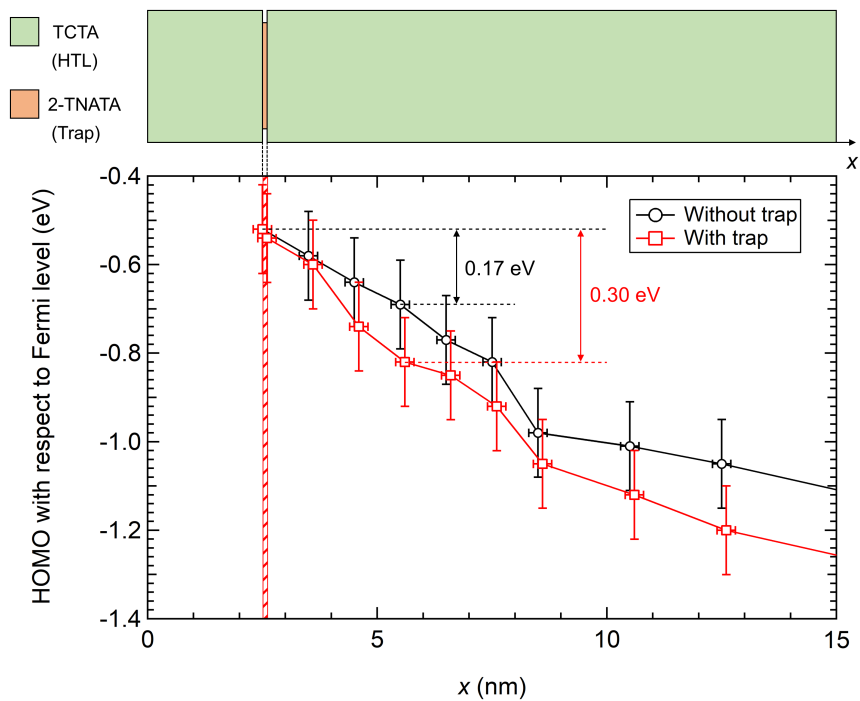

**FIG. S4. The HOMO energy level of the HTL without (black circle) and with (red square) traps measured with respect to the Fermi level.**

To investigate the trap-induced band bending of the HTL, we show in **Fig. S4** the shift in the HOMO energy level with respect to the Fermi level for the HTL with and without traps, given as a function of the distance to the anode ($x$). The UPS spectra for the samples were traced *in-situ* during the growth of the HTL layer with an increasing thickness up to ~15 nm. For the trap-containing HTL, traps were inserted in proximity to the ITO anode at a distance of 2.5 nm. The trap-containing HTL (red square) exhibits a larger downward shift in the HOMO energy level at $x = 5$ nm by $0.30 \pm 0.10$ eV vs. $0.17 \pm 0.10$ eV for the trap-free HTL (black circle) with respect to the respective HOMO energy levels at $x = 2.5$ nm. This measurement supports our argument that the localized holes by the traps near the anode lead to the stronger band bending of the HTL compared to that of the trap-free HTL.

## Sec. 4. The sole effect of the trap position on the electric field

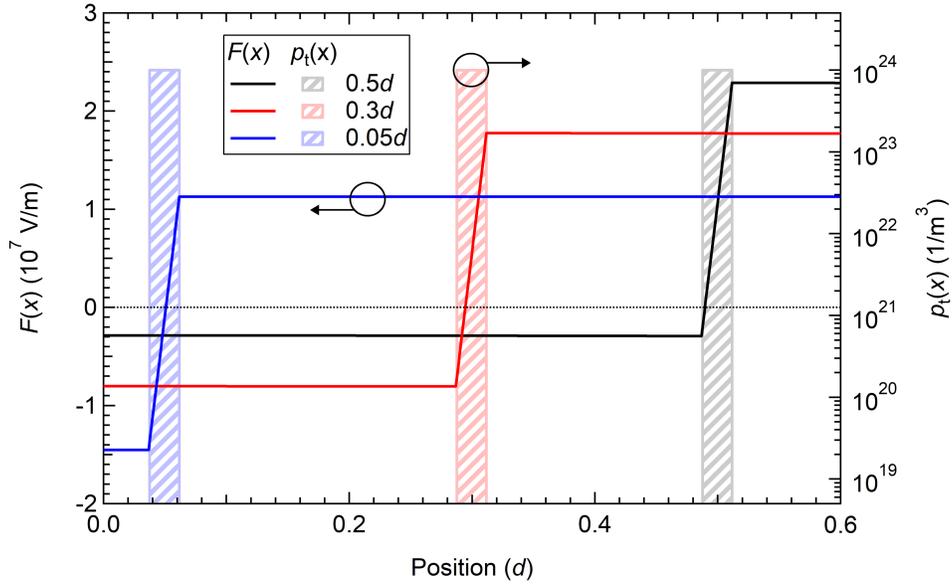

**FIG. S5.** The electric field profile, $F(x)$, for the three HODs containing traps at the respective position of $x_t$ = 0.05$d$, 0.3$d$ and 0.5$d$. The profile was obtained by solving the Poisson's equation (Eq. (S1)) with $p(x) = 0$, $p_t(x) = 10^{24}$ m$^{-3}$ and $V(0) = 2$ V for the given $x_t$ value.

To investigate the change in the electric field profile, $F(x)$, exclusively due to the trap position at $x = x_t$, we solved the Poisson's equation for $F(x)$ by varying $x_t$ [**Eq. (S1)**] while the other essential parameters fixed as constants. For example, we assume that (i) free holes are absent in the HODs, i.e., $p(x) = 0$, (ii) the trapped hole density remains constant at $p_t(x_t) = 10^{24}$ m$^{-3}$ for any given $x_t$, and (iii) the driving voltage, $V_a$, is set to yield a constant electric potential at the anode/HTL interface as $V(x = 0) = 2$ V. **Fig. S5** shows the $F(x)$ for three selected HODs with traps inserted at the position of $x_t$ = 0.05$d$, 0.3$d$, and 0.5$d$. The trends discussed in the main text can be more clearly found with decreasing $x_t$ – that is – the magnitude of the electric field exerted by $p_t(x_t)$ towards the cathode ($F > 0$ in the region of $x_t \leq x \leq d$) decreases while that of the electric field towards the anode ($F < 0$) becomes progressively intensified or *more negative*.